\documentclass[12pt]{article}
\pdfoutput=1
\newlength{\mytopmargin}
\newlength{\myleftmargin}
\newtheorem{theorem}{Theorem}[section]
\newtheorem{proposition}[theorem]{Proposition}

\usepackage{amsmath,amsfonts,amssymb}

\usepackage{graphicx}

\begin{document}
\title{One-component plasma on a spherical annulus and a random matrix ensemble}
\author{Jonit Fischmann${}^*$ and Peter J. Forrester${}^\dagger$}
\date{}
\maketitle

\noindent
\thanks{\small
${}^*$ School of Mathematical Sciences, Queen Mary University of London,
 London E1 4NS, UK email: { j.fischmann@qmul.ac.uk}\\
${}^\dagger$ Department of Mathematics and Statistics,
The University of Melbourne,
Victoria 3010, Australia email: { P.Forrester@ms.unimelb.edu.au}
}

\begin{abstract}
The two-dimensional one-component plasma at the special coupling $\beta = 2$ is known to be exactly solvable,
for its free energy and all of its correlations, on a variety of surfaces and with various boundary conditions.
Here we study this system confined to a spherical annulus with soft wall boundary conditions,
paying special attention to the resulting asymptotic forms from the viewpoint of expected general properties of the two-dimensional plasma.
Our study is motivated by the realization of the Boltzmann factor for the plasma system with $\beta = 2$, after stereographic projection from the
sphere to the complex plane, by a certain random matrix ensemble constructed out of complex Gaussian and Haar distributed
unitary matrices.
\noindent   \end{abstract}

\section{Introduction}
The two-dimensional one-component plasma is an equilibrium statistical mechanical system consisting of $N$ mobile particles, each of charge $+1$, and a smeared out neutralizing background. The particles are confined to a two-dimensional surface, and the charge densities (both point and continuous) interact through the solution of the two-dimensional Poisson equation on the surface.

Although it is defined as a classical system, the two-dimensional one-component plasma in the case that the surface is of constant curvature is also known in quantum many body physics. This is due to its relevance to the fractional quantum Hall effect. Thus it turns out that the Boltzmann factor for the plasma system at inverse temperature $\beta = 2 \nu$, $\nu$ an odd integer, is equal to the absolute value squared of the Laughlin trial wave function for the fractional quantum Hall effect at filling fraction $1/\nu$ \cite{La83, Ha83x, Du92a}. In the case $\nu = 1$ and thus $\beta = 2$ the corresponding trial wave function is in fact the exact wave function for non-interacting spinless fermions with constant perpendicular magnetic field.

It has been known for some time that there is also an analogy between the two-dimensional one-component plasma confined to a disk in the plane, and the complex Ginibre random matrix ensemble  \cite{AJ81}. The latter is specified as the eigenvalue probability density function (PDF) for $N \times N$ complex Gaussian matrices, where each element is independently distributed as a standard complex Gaussian. In terms of the notation $z_j = x_j + i y_j$, $x_j,y_j \in \mathbb R$, it has the explicit form
\begin{equation}\label{Z}
\prod_{l=1}^N e^{-|z_l|^2} \prod_{1 \le j < k \le N} |z_k - z_j|^2,
\end{equation}
up to proportionality. If the extra condition that $|z_l| \le \sqrt{N}$ is imposed, then (\ref{Z}) is
proportional to the Boltzmann factor for the one-component plasma at coupling $\beta = 2$,
confined to a disk of radius $R = \sqrt{N}$. Without this constraint, the eigenvalues are to leading order still confined to a disk in this radius (an example of the circular law \cite{Gi84,Ba97,GT10,TV08}).

More recently, an analogy between two other random matrix ensembles and the one-component plasma confined to the other homogeneous constant curvature two-dimensional surfaces --- namely the sphere and pseudosphere --- has been specified. Thus in \cite{Kr06} it was shown that the eigenvalue PDF for random matrices $A^{-1} B$, where $A$ and $B$ are independent complex Ginibre matrices, coincides with the Boltzmann factor for the one-component plasma at $\beta = 2$ on the sphere, after a stereographic projection of the latter. And in \cite{FK09} it was shown that the eigenvalue PDF of truncations of unitary random matrices \cite{ZS99} has the same form as the Boltzmann factor for the one-component plasma on the pseudosphere at $\beta = 2$, after projection of the latter onto the Poincar\'e disk. These examples of the one-component plasma had earlier been identified as exactly solvable two-dimensional statistical mechanical systems \cite{Ca81,FJM92,JT98}. As an aside we mention that the one-component plasma confined to a surface of non-constant curvature---Flamm's paraboloid which occurs as the spatial part of the Schwarzschild metric from general relativity in two-dimensions---has recently been shown to also be exactly solvable at
$\beta = 2$ \cite{FT08}, although as yet no random matrix analogy has been found.

A topic of much current interest in random matrix theory is ensembles formed from the product
$U Y^{1/2}$, where $U$ is a unitary random matrix and $Y$ is positive definite
\cite{HL00, FW08, Gu09, Bo10}. The motivation behind our work is to relate, for a particular class of random matrices $Y$ generalizing the ensemble
$A^{-1}B$, an eigenvalue PDF obtained in this setting to the two-dimensional one-component plasma at $\beta = 2$ confined to a spherical annulus. The system is exactly solvable, being an example of a determinantal point process. Moreover, we will see that the asymptotic forms of the partition function, one and two point correlations, and the distribution of a general axially symmetric linear statistic all illustrate physical properties of the point process which are expected to hold for the plasma system in the same geometry but with $\beta > 0$ \cite{Fo98a}.

In Section 2 the Boltzmann factor for the one-component plasma confined to a spherical annulus is calculated, as is its form upon
a stereographic projection. In the case $\beta = 2$, and with the area of the spherical caps outside the spherical annulus certain rational fractions
of the area of the sphere, a realization of the projected functional form of the Boltzmann factor as the eigenvalue PDF of a random matrix
ensemble is given in Section 3. In Sections 4 and 5 the plasma system at $\beta = 2$ is studied as an exactly solvable statistical mechanical model,
and the corresponding large $N$ asymptotic forms are computed and used according to the final sentence of the above paragraph.

\section{The plasma system}
\setcounter{equation}{0}
Consider a sphere $S$ of radius $R$, and let  $0  \le \theta \le  \pi$ refer to the usual azimuthal angle,
and  $0 \le \phi \le 2\pi$ refer to the polar angle. For two points
$(\theta,\phi)$ and $(\theta',\phi')$ on the sphere, let $\alpha$ refer to their relative angle
when considered as vectors in $\mathbb R^3$.
We know that the solution of the charge neutral
Poisson equation
$$
\nabla^2_{\theta,\phi} \Phi = -2 \pi \delta_S((\theta,\phi),(\theta',\phi')) + {1 \over 2 R^2}
$$
(the sphere being a compact surface, charge neutrality is a necessary condition for existence
of a solution), where $ \delta_S((\theta,\phi),(\theta',\phi'))$ is the delta function on the sphere, is then given by \cite{Ca81}
\begin{equation}\label{P}
 \Phi ((\theta,\phi),(\theta',\phi')) = - \log(2 R \sin(\alpha/2)).
 \end{equation}
 Introducing the Cayley-Klein parameters,
 \begin{equation}\label{P1}
 u := \cos(\theta/2) e^{i \phi/2}, \qquad v := -i \sin(\theta/2) e^{-i \phi/2}
 \end{equation}
 we know (see e.g.~\cite[eq.~(15.108)]{Fo10}) that (\ref{P}) can be rewritten
\begin{equation}\label{P2}
 \Phi ((\theta,\phi),(\theta',\phi')) = - \log(2 R |u'v - u v'|).
 \end{equation}

 Let us mark two circles on the sphere corresponding to the azimuthal angles $\theta_Q$ and $\pi - \theta_q$, with
 $0 < \theta_Q < \pi - \theta_q < \pi$. The surface of the sphere between these circles defines
 a spherical annulus. Let $A_{[0,\theta_Q]}$ denote the area of the spherical cap above
 $\theta_Q$ and thus including the north pole, and let $A_{[\pi - \theta_q,\pi]}$ denote the
 area of the spherical cap below $\pi - \theta_q$ and thus including the south pole. We parametrize
 $\theta_Q$ and $\theta_q$ by introducing $Q$ and $q$ such that
 \begin{equation}\label{A}
{ A_{[0,\theta_Q]} \over 4 \pi R^2} = {Q \over 1 + q + Q}, \qquad
{ A_{[\pi-\theta_q,\pi]} \over 4 \pi R^2} = {q \over 1 + q + Q}.
\end{equation}

The plasma is specified by requiring that
 within the spherical annulus  there be $N$ mobile
 particles of charge $+1$ and a uniform neutralizing background. Both the discrete
 and continuous charges are to interact via the potential (\ref{P}). It follows from
 (\ref{A}) that the area of the annulus $A_{[\theta_Q, \pi - \theta_q]}$  is such that the
 uniform neutralizing background charge density is equal to
 \begin{equation}\label{A1}
 -{N \over 4 \pi R^2} (1 + Q + q)=: - \rho_b.
 \end{equation}

 We would like to compute the potential energy $V(\theta')$ of the interaction of a particle at
 $(\theta',\phi')$ in the spherical annulus, and the neutralizing background.
 For this purpose
 we extend the background to have uniform charge density $-\rho_b$ throughout the sphere.
 To compensate, we must impose a uniform charge density $\rho_b$ in the spherical
 caps above $\theta_Q$ and  below $\pi - \theta_q$.
 We can now proceed to compute the sought
 potential. Throughout we will ignore the $2R$ factor in the logarithm of (\ref{P2}): by
 charge neutrality, we can check that it must contribute a factor $(2 R)^{N \beta/2}$
 to the Boltzmann factor.

 \begin{proposition}\label{p1}
 We have that
 \begin{equation}\label{V}
 V(\theta') =  C_N  -NQ \log \sin {\theta' \over 2} -
  N q \log \cos {\theta' \over 2},
  \end{equation}
  where
 \begin{equation}\label{V1}
 C_N :=  -{N \over 2} + {N \over 2} (1 + q) \log {1 + q \over 1 + Q + q} +
  {N \over 2} (1 + Q) \log {1 + Q \over 1 + Q + q}.
   \end{equation}
  \end{proposition}

  \noindent {Proof.} \quad
 The potential of the interaction of a particle with the uniform background covering all
 the sphere is independent of the location of the particle. Choosing this location to be
 the north pole, we see from (\ref{P2}) and the fact that on the surface of a sphere
 $dS = R^2 \sin \theta \, d\theta d \phi$ that the corresponding potential energy is
 \begin{equation}\label{S}
 \rho_b R^2 \int_0^\pi \sin \theta \Big ( \log \sin {\theta \over 2} \Big ) d \theta \int_0^{2 \pi} d \phi.
\end{equation}
 Using the integral evaluation
  \begin{equation}\label{I}
 \int_0^t \sin x \Big ( \log \sin {x \over 2} \Big ) \, dx =
 \Big ( -1 + 2 \log \sin {t \over 2} \Big ) \Big ( \sin {t \over 2} \Big )^2
 \end{equation}
 with $t = \pi$ we see that (\ref{S}) simplifies to
  \begin{equation}\label{S1}
  - {N \over 2} (1 + Q + q).
 \end{equation}

 Consider next the potential between a particle and the charge density $\rho_b$ in the
 spherical cap above $\theta_Q$. This is equal to
 \begin{equation}\label{S2}
 - \rho_b R^2 \int_0^{\theta_Q} d \theta \, \sin \theta \int_0^{2 \pi} d \phi \,
 \log | u' v - u v'|.
 \end{equation}
 Simple manipulation gives
 \begin{equation}\label{S3}
 \log |u' v - u v'| = \log \cos {\theta \over 2} + \log \sin {\theta' \over 2} +
 \log \Big | 1 - {\tan \theta/2 \over \tan \theta'/2} e^{- i (\phi - \phi')} \Big |.
 \end{equation}
 Note that the ratio
  of tan functions has magnitude less than one.
 Substituting into (\ref{S2}), this latter fact implies the third term in
  (\ref{S3}) does not contribute since the integral over $\phi$ vanishes, and hence (\ref{S2}) reduces to
\begin{equation}\label{S4}
- \rho_b R^2 (2 \pi) \int_0^{\theta_Q} \sin \theta \Big ( \log \cos {\theta \over 2} + \log \sin {\theta' \over 2} \Big ) \, d \theta.
\end{equation}
This simplifies by noting from (\ref{A}) that
\begin{equation}\label{S5}
R^2 (2 \pi) \int_0^{\theta_Q} \sin \theta  \, d \theta = (4 \pi R^2) {Q \over 1 + Q + q},
\end{equation}
while use of (\ref{I}) shows that
\begin{align}\label{S6}
 \int_0^{\theta_Q} \sin \theta  \log \cos {\theta \over 2} \, d \theta & =
 - \sin^2 {\theta_Q \over 2} - \cos^2  {\theta_Q \over 2} \log \cos^2  {\theta_Q \over 2} \nonumber \\
 & = - {Q \over 1 + Q + q} - {1 + q \over 1 + Q + q} \log {1 + q \over  1 + Q + q},
 \end{align}
where the second equality follows by making use of (\ref{S5}). Substituting (\ref{S5}) and
(\ref{S6}) in   (\ref{S4}) we conclude that the potential between a particle and the charge density $\rho_b$ in the
 spherical cap above $\theta_Q$ is equal to
  \begin{equation}\label{S1a}
  -  N Q \log \sin {\theta' \over 2}  + {N \over 2} \Big ( Q + (1 + q) \log {1 + q \over 1 + Q + q} \Big ).
 \end{equation}

 Replacing $q$ by $Q$ and $\theta'$ by $\pi - \theta'$ gives that the potential between a particle
 and the charge density $\rho_b$ in the spherical cap below $\pi - \theta_q$ is
 equal to
 \begin{equation}\label{S1b}
  -  N q \log \cos {\theta' \over 2}  + {N \over 2} \Big ( q + (1 + Q) \log {1 + Q \over 1 + Q + q} \Big ).
 \end{equation}
 Adding together (\ref{S1}), (\ref{S1a}) and (\ref{S1b}) gives (\ref{V}). \hfill $\square$\\

 Note that an equivalent viewpoint on the result (\ref{V}) is that the potential
 $$
  -NQ \log \sin {\theta \over 2} -
  N q \log \cos {\theta \over 2}
  $$
  results from charges $NQ$ and $Nq$ at the north and south poles respectively.
  With $\alpha_j$ denoting the angle between a point $(\theta, \phi)$ on the sphere, and
  another  point $(\theta_j, \phi_j)$, a related question is to seek the background charge density which gives rise to the potential
  $$
  - N \sum_{j=1}^p q_j \log \sin (\alpha_j/2).
  $$
 In a disk geometry, the analogous question has recently been addressed in \cite{BH09}.

 We turn our attention next to the computation of the potential for the interaction of the
 background with itself.
  \begin{proposition}
 The background-background potential is equal to
 \begin{align}\label{S1c}
 & {N^2 \over 4} - {N^2 \over 4} (1 + q) \log {1 + q \over 1 + Q + q}
 - {N^2 \over 4} (1 + Q) \log  {1 + Q \over 1 + Q + q}  \nonumber \\
 & + {N^2 \over 4} \Big ( - (Q + q) + Q \log {1 \over 1 + Q + q} +
 Q (1 + Q) \log (1 + Q) \log (1 + Q) - Q^2 \log q  \nonumber \\
 & \qquad \qquad
 q \log {1 \over 1 + Q + q} +
 q (1 + q) \log (1 + q) \log (1 + q) - q^2 \log Q \Big ).
 \end{align}
 \end{proposition}

 \noindent {Proof.} \quad The  background-background potential is given in terms of
 the particle background potential $V(\theta)$ according to
 \begin{equation}\label{A1}
 - {1 \over 2} \rho_b (2 \pi R^2) \int_{\theta_Q}^{\pi - \theta_q}
 \sin \theta \, V(\theta) \, d \theta.
 \end{equation}
 Substituting (\ref{V}) and performing the first of the resulting integrals gives
 $$
 -{N \over 2} C_N + {N \over 4} (1 + Q + q)
 \int_{\theta_Q}^{\pi - \theta_q} \sin \theta \Big (
 N Q \log \sin {\theta \over 2} + N q \log \cos {\theta \over 2} \Big ) \, d \theta.
 $$
 The integrals can be performed using (\ref{I}) and further reduced as in the second
 equality of (\ref{S6}), with the result being (\ref{S1c}).
 \hfill $\square$\\

 The total potential energy $U$ of the plasma system consists of the particle-particle,
 particle-background, and background-background interactions. It therefore follows from
 (\ref{P2}), (\ref{V}), (\ref{S1c}) and the remark above Proposition \ref{p1}
 that the Boltzmann factor $e^{-\beta U}$
 for the plasma system is equal to
 \begin{equation}\label{C1}
 \Big ( {1 \over 2 R} \Big )^{N \beta/2} e^{-\beta K_N}
 \prod_{l=1}^N | v_l|^{\beta QN} |u_l|^{\beta q N}
 \prod_{1 \le j < k \le N} | u_k v_j - u_j v_k|^\beta,
 \end{equation}
 where
 \begin{align}\label{C2}
 K_N :=  & {N^2 \over 4} \Big (
 - (1 + Q + q) + 2 (1 + Q + q) \log {1 \over  1 + Q + q} + (1 + q)^2 \log (1 + q) \nonumber \\
 & \qquad + (1 + Q)^2 \log (1 + Q) - Q^2 \log q - q^2 \log Q \Big ).
 \end{align}
 By construction the particles are restricted to the spherical annulus. However, as we will see,
 the analogy between the Boltzmann factor and the plasma and the eigenvalue PDF for a certain
 random matrix ensemble requires that this constraint be relaxed. Nonetheless, we will find that
 up to terms which vanish as a Gaussian, the support of the eigenvalue PDF is still the spherical
 annulus. It should be mentioned that this analogy assumes a particular transformation of the eigenvalues,
 which start out as points in the complex plane. The mapping from a point
 $z = x + i y$ in the
 complex plane, to a point $(\theta, \phi)$ on the sphere, is carried out by the stereographic
 projection
 \begin{equation}\label{zz}
 z = 2R e^{i \phi} \tan {\theta \over 2}.
 \end{equation}
 We know from e.g.~\cite[eqns.~(15.126), (15.127)]{Fo10} that then
 $$
 2 R  | u' v - u v'| = \cos {\theta \over 2} | z - z'| \cos {\theta' \over 2}, \qquad
 dS = {1 \over (1 + |z|^2/4 R^2)^2} dx dy.
 $$
 Consequently, with $\tilde{z} := z/(2R)$,
 \begin{eqnarray}\label{pz}
 &&
 \prod_{l=1}^N | v_l|^{\beta QN} |u_l|^{\beta q N}
 \prod_{1 \le j < k \le N} | u_k v_j - u_j v_k|^\beta dS_1 \cdots dS_N
  = \prod_{l=1}^N \Big ( {|\tilde{z}_l|^2 \over 1 + |\tilde{z}_l|^2} \Big )^{ \beta Q N/2} \nonumber \\
 &&
 \quad \times
 {1 \over (1 + |\tilde{z}_l|^2 )^{ \beta q N/2 + 2 + \beta (N-1)/2}}
 \prod_{1 \le j < k \le N} | \tilde{z}_j - \tilde{z}_k |^\beta d \vec{r}_1 \cdots  d \vec{r}_N.
 \end{eqnarray}

 We remark that the spherical annulus bounded between the azimuthal angles
 $\theta_Q$ and $\theta_{\pi - \theta_q}$ maps, under the stereographic
 projection (\ref{zz}),  to a planar annulus with radii $r_Q$ and $r_q$. Making use
 of (\ref{zz}), together with  (\ref{A}) it follows that
 \begin{equation}\label{rr}
 \Big ({r_Q \over 2 R} \Big )^2 = {Q \over 1 + q} =: \tilde{r}_Q^2, \qquad
  \Big ({r_q \over 2 R} \Big )^2 = {1 + Q \over q} =: \tilde{r}_q^2.
  \end{equation}

 \section{Analogy with a random matrix ensemble}
 \setcounter{equation}{0}

 Let $A$ and $B$ be $N\times N$ random matrices, with entries independently  chosen as standard complex Gaussians. It was shown by Krishnapur \cite{Kr06} that the eigenvalue PDF of $A^{-1}B$ is, up to normalisation, given by the RHS of (\ref{pz}) with $\beta=2$, $q=Q=0$. In this section a more general random matrix realization of (\ref{pz}) will be given, applying for $\beta=2$ and arbitrary $qN$, $QN\in\mathbb{Z}_{\geq0}$.

 To achieve this, two results from random matrix theory must be combined. In relation to the first, with an $n\times M$, $n\geq M$ , standard complex Gaussian matrix $a$, set $A=a^\dagger a$ to form a so-called complex Wishart matrix (see e.g. \cite[Ch. 3]{Fo10}). Let $X$ be an $M\times N$, $N\geq M$, standard complex Gaussian matrix, then set $Y=A^{-1/2}X$. We know from \cite{GN97} that, up to normalization, the element joint probability density function of $Y$ is given by
 $$
 \frac{1}{{\rm det}({\mathbb I}+Y^\dagger Y)^{n+N}}.
 $$

 In relation to the second of the results, suppose $W$ is an $M\times N$ random matrix with element PDF of the form $g(WW^\dagger)$. Also, let $U$ be an $M\times M$ unitary random matrix chosen with Haar measure. Then we know from \cite{FBKSZ11} that with $N\geq M$ and up to normalization the PDF of $G=U(W W^\dagger)^{1/2}$ is given by
 $$
( {\rm det} G^\dagger G)^{N-M}g(G^\dagger G).
 $$

Let us choose $W$ in the second result according to $Y$ as specified in the first. This shows that the element PDF of $G=U(Y^\dagger Y)^{1/2}$ is proportional to
\begin{equation} \label{GG}
( {\rm det} G^\dagger G)^{N-M} \frac{1}{{\rm det}({\mathbb I}+G^\dagger G)^{n+N}}.
 \end{equation}
 The explicit value of the proportionality constant can readily be calculated.

 \begin{proposition}\label{pp1}
 Let (\ref{GG}) when multiplied by $1/\mathcal{N}$ be correctly normalized. Then we have
 \begin{equation} \label{GGN}
 \mathcal{N}=\pi^{M^2}\prod_{j=0}^{M-1}\frac{\Gamma(N-M+1+j)\Gamma(n-M+1+j)}{\Gamma(n+N-M+1+j)\Gamma(1+j)}.
 \end{equation}
 For this to be well defined we require $N\geq M$ and $n \ge M$.
\end{proposition}
  \noindent {Proof.} \quad
With $C=GG^\dagger$ and the eigenvalues of $C$ written $\{\lambda_j\}_{j=1,\ldots,M}$ we know that
\begin{equation}\label{Gl}
(dG)=\tilde{c}\prod_{1\leq j<k \leq M} (\lambda_k-\lambda_j)^2 d\lambda_1\ldots \lambda_M.
\end{equation}
Here $\tilde{c}$ is independent of the eigenvalues and $(dG)$ denotes the product of differentials of the independent real and imaginary parts. To determine $\tilde{c}$, suppose temporarily that $G$ is a standard complex normal random matrix so that it has PDF
\begin{equation}\label{Gl1}
\pi^{-M^2}e^{-{\rm Tr}\,G^\dagger G}=\pi^{-M^2}e^{-\sum_{j=1}^M \lambda_j}.
\end{equation}
Converting now to the corresponding measures on both sides using $(\ref{Gl})$ then integrating shows
\begin{equation}\label{Gl2}
1=\pi^{-M^2}\tilde{c}\int_0^\infty d \lambda_1 \cdots \int_0^\infty d\lambda_M\,  e^{-\sum_{j=1}^M \lambda_j}\prod_{1\leq j<k \leq M} (\lambda_k-\lambda_j)^2.
\end{equation}
Evaluation of the integral (see e.g. \cite[Prop. 4.7.3]{Fo10}) now gives
\begin{equation}\label{ct}
\tilde{c}=\frac{\pi^{M^2}}{\prod_{j=0}^{M-1}\Gamma(1+j)\Gamma(2+j)}.
\end{equation}

With $\tilde{c}$ determined, we can proceed to evaluate $\mathcal{N}$ using an analogous strategy. Thus after multiplying (\ref{GG}) by $1/\mathcal{N}$ so that it is normalized from the analogue of (\ref{Gl1}) by introducing the eigenvalues of $G^\dagger G$. We then use (\ref{Gl}) to convert that equation into an equality of measures. Integrating both sides, then changing variables $\lambda_j=t_j/(1-t_j)$ $(j=1,\ldots, M)$ on the RHS we obtain
$$
1=\frac{\tilde{c}}{\mathcal{N}}\int_0^1 dt_1 \ldots \int_0^1dt_M\prod_{j=1}^Mt_j^{N-M}(1-t_j)^{n-M}\prod_{1\leq j <k\leq M} (t_k-t_j)^2.
$$
The multi-dimensional integral herein is a special case of the Selberg integral (see e.g. \cite{FW08}, \cite[Ch. 4]{Fo10}). It's evaluation as a product of gamma functions together with (\ref{ct}) gives (\ref{GGN})
\hfill $\square$\\

We seek the eigenvalue PDF implied by the element PDF $(\ref{GG})$, normalized according to Proposition \ref{pp1}. Of course $G$ as defined above (\ref{GG}) is non-Hermitian, and the eigenvalues will lie in the complex plane (for reviews of aspects of the rich mathematical physics associated with this setting see \cite{FS03}, \cite{Ve09}, \cite{KS09}, \cite[Ch. 15.]{Fo10}). We will see that the eigenvalue PDF can be identified with the RHS of (\ref{pz}) in the case $\beta=2$, $N=M$ and $qN$, $QN\in \mathbb{Z}_{\geq 0}$ arbitrary.

\begin{proposition}
Let $G$ be an $M\times M$ matrix with element PDF (\ref{GG}) and normalized by (\ref{GGN}). The corresponding eigenvalue PDF is given by
\begin{equation}\label{C1}
\frac{1}{\mathcal{C}}\prod_{j=1}^{M}\frac{|z_j|^{2(N-M)}}{(1+|z_j|^2)^{n+N-M+1}}\prod_{1\leq j < k \leq M}|z_k-z_j|^2,
\end{equation}
where
\begin{equation}\label{C2}
\mathcal{C}=M!\pi^M\prod_{j=0}^{M-1}\frac{\Gamma(N-M+1+j)\Gamma(n-M+1+j)}{\Gamma(n+N-M+1)}.
\end{equation}
\end{proposition}

  \noindent {Proof.} \quad
We follow \cite{HKPV08} (see also \cite[Prop. 15.6.1]{Fo10}). The first step is to introduce the complex Schur decomposition by writing $G=URU^\dagger$ where $U$ is an $M\times M$ unitary matrix and $R=\Lambda+T$, with $\Lambda={\rm diag}(z_1,\ldots,z_M)$ the diagonal matrix of eigenvalues and $T$ strictly upper triangular.

To make the decomposition unique, we must order the eigenvalues (for example, according to their modulus) and choose $U$ from the right coset of the unitary group $\mathcal{U}[M]:=U(M)/U_d(M)$, where $U_d(M)$ denotes the set of diagonal $M\times M$ unitary matrices. The corresponding volume form is given by $(\mathcal{U}^\dagger d \mathcal{U})$. For later reference we note that (see e.g. \cite[eq. (3.23)]{Fo10})
\begin{equation}\label{UU}
\int(\mathcal{U}^\dagger d \mathcal{U})=\frac{\pi^{M(M-1)/2}}{\prod_{j=0}^{M-1}\Gamma(j+1)}.
\end{equation}

We know that the change of variables formula from $G$ to $U$ and $R$ is (see e.g. \cite[Prop. 15.1.1]{Fo10})
$$
(dG)=\prod_{1\leq j<k\leq M}|z_k-z_j|^2\prod_{j=1}^Mdx_jdy_j\prod_{1\leq j<k\leq M}dT_{jk}^r
dT_{jk}^i(\mathcal{U}^\dagger d \mathcal{U}),$$
where $z_j=x_j+iy_j$ and $dT_{jk}=dT_{jk}^r+idT_{jk}^i.$ To obtain the eigenvalue PDF we must multiply this by (\ref{GG}), together with its normalization, and itegrate over $\mathcal{U}$ and $T$. Thus the eigenvalue PDF of $G$ is equal to
\begin{eqnarray}\label{6.7}
\frac{1}{\mathcal{N}}\left(\int(\mathcal{U}^\dagger d \mathcal{U}) \right)\prod_{j=1}^M|z_j|^{2(N-M)}\prod_{1\leq j<k\leq M}|z_k-z_j|^2 \notag \\
\times \int\frac{1}{{\rm det}({\bf 1}+R^\dagger R)^{n+N}}\prod_{1\leq j<k \leq M} dT_{jk}^rdT_{jk}^i.
\end{eqnarray}

Let $\mathbf{v}_{m-1}$ be an $(m-1)\times 1$ complex vector, and set
\begin{equation}\label{6.7a}
c_{m,p}=\int\frac{(d \mathbf{v}_{m-1})}{(1+\mathbf{v}_{m-1}^\dagger \mathbf{v}_{m-1})^p}=\pi^{m-1}\frac{\Gamma(p-m+1)}{\Gamma(p)}.
\end{equation}
Also, after writing $R=R_m$ to indicate the size of $R$, set
$$
I_{m,p}(z_1,\ldots, z_m):=\int\frac{1}{{\rm det}(\mathbb{I}+R^\dagger_mR_m)^p}\prod_{1\leq j<k\leq m} dT_{jk}^rdT_{jk}^i.
$$
Then we know from \cite[eq. (15.138)]{Fo10} that
$$
I_{m,p}(z_1,\ldots, z_m)=\frac{c_{m,p}}{(1+|z_m|^{p-m+1})}I_{m-1,p-1}(z_1,\ldots,z_{m-1}).
$$
This allows the final integral to be evaluated as
$$
\prod_{l=0}^{M-1}\frac{c_{M-l,n+N-l}}{(1+|z_{M-l}|^2)^{n+N-M+1}}.
$$
Substituting in (\ref{6.7}) and simplifying using (\ref{6.7a}), (\ref{UU}) and (\ref{GGN}) gives (\ref{C1}). In the normalization (\ref{C2}),
the ordering on the eigenvalues has been relaxed.
\hfill $\square$\\

Comparing (\ref{C1}) with the RHS of (\ref{pz}) we see that they agree if in the latter we set $\beta=2$, $N=M$ and $$
QN=N-M, \hspace{1cm} qN=n-M.
$$
In Figure 1 we show numerically generated eigenvalues corresponding to the choice $Q=q=1$,
stereographically projected onto the sphere. This illustrates the eigenvalue density being, to leading
order, uniform within the spherical annulus, and zero outside.

\begin{figure}
\label{fig:}
\begin{center}
\includegraphics[scale=0.6]{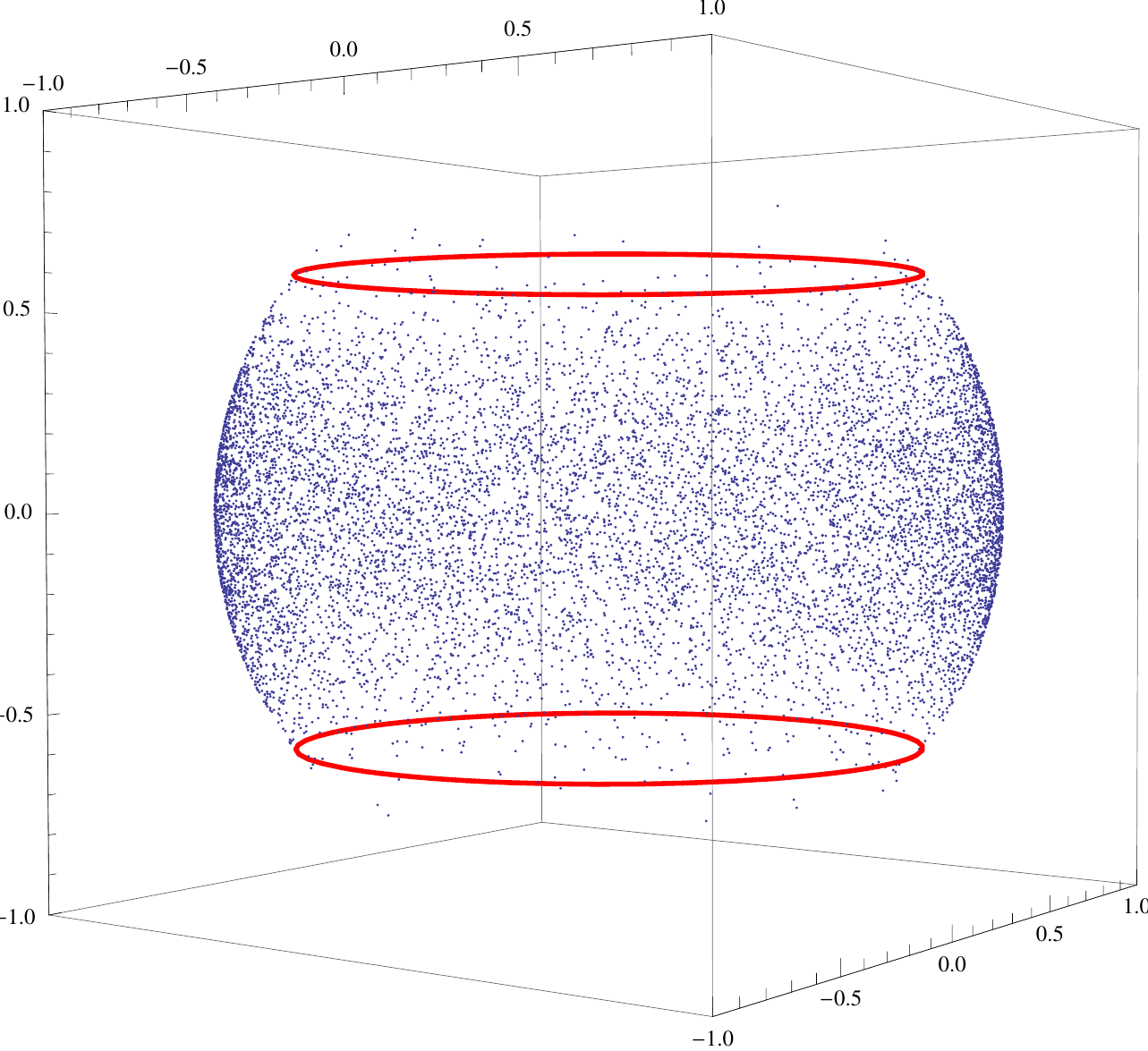}
\end{center}
\caption{Stereograpically projected eigenvalues of matrices with element PDF (\ref{GG}) and
eigenvalue PDF (\ref{C1}) in the case $n=N=20$, $M=10$, repeated 1,000 times. The marked circles
are the theoretical boundaries of support for $n=N=2M$ and $M \to \infty$. The sphere has been scaled to have
radius 1.}
\end{figure}

 \section{Free energy}\label{Section4}
 \setcounter{equation}{0}
 Let us return now to the plasma interpretation of (\ref{C1}).
 A primary quantity of interest is then the large $N$ form of the dimensionless free energy,
 \begin{equation}\label{FZ}
 \beta F_N = - \log Z_N(\beta),
 \end{equation}
where $Z_N(\beta)$ is the partition function
\begin{eqnarray}\label{FZ1}
Z_N(\beta) = {1 \over N!}  \Big ( {1 \over 2 R} \Big )^{N \beta/2} e^{-\beta K_N}
\prod_{l=1}^N R^2 \int_0^{2 \pi} d \phi_l \int_0^\pi d \theta_l \,
| v_l|^{\beta QN} |u_l|^{\beta q N}
 \prod_{1 \le j < k \le N} | u_k v_j - u_j v_k|^\beta.
 \end{eqnarray}
 We know from studies relating the two-dimensional Coulomb gas to the Gaussian
 free field \cite{JMP94} that the large $N$ expansion of $\log Z_N$ should be of the form
 \begin{equation}\label{FZ2}
 \log Z_N(\beta) \: \sim \: A_\beta N + B_\beta N^{1/2} + {\chi \over 12} \log N + \cdots.
 \end{equation}
 Here $-A_\beta$ is the dimensionless free energy per particle, $-B_\beta$ is the
 dimensionless  surface tension,
 and $\chi$ denotes the Euler characteristic of the surface (explicitly $\chi = 1$ for a disk, $\chi = 2$ for a sphere, $\chi = 0$ for an annulus).

 The fact that the leading term in (\ref{FZ2}) is proportional to $N$ follows from the proof of the existence of the
 thermodynamic limit for jellium by Lieb and Narhofer \cite{LN75}. This term is a bulk quantity,
 and so is independent of the geometry. We know from \cite{AJ81} in the case of a disk that
 for $\beta = 2$
 \begin{equation}\label{Fa}
 A_2 = -{1 \over 2} \log {\rho_b \over 2 \pi^2}.
 \end{equation}
 That this is indeed independent of the geometry has been illustrated by exact calculation in the
 case of the sphere \cite{Ca81}, for example. Again from exact calculations in the case of the
 disk at $\beta = 2$, the exact form of $B_2$ is known. It is expected to be dependent only on the
 length of the boundary, and exact calculation in the case of semi-periodic boundary conditions
\cite{CFS83}  illustrates this. In the case of soft wall boundary conditions, when the mobile particles
are not confined to the region initially assumed in the computation of the Boltzmann factor, it
has been observed in exact calculations \cite{DGIL94} that $B_2=0$. As we are interested
in the case of soft wall boundary conditions, we thus expect that
\begin{equation}\label{Fb}
 B_2 = 0.
 \end{equation}
 Hence the formula (\ref{FZ2}) predicts that for the plasma confined to the soft wall spherical
 annulus
\begin{equation}\label{Fc}
 \log Z_N(2) \: \sim \:   -{N \over 2} \log {\rho_b \over 2 \pi^2} + {\rm O}(1).
 \end{equation}
 Starting with (\ref{FZ1}), standard integration methods (see e.g.~\cite[\S 15.3]{Fo10}) verify
 (\ref{Fc}), and furthermore allow us to explicitly compute the term ${\rm O}(1)$.

 \begin{proposition}
 With $\beta = 2$,  the asymptotic expansion of (\ref{FZ1}) for large
 $N$ reads
 \begin{equation}\label{Fd}
  \log Z_N(2) \: \sim \:   -{N \over 2} \log {\rho_b \over 2 \pi^2} +
  {1 \over 12} \log {Q \over 1 + Q} + {1 \over 12} \log {q \over 1 + q} +
  {\rm O}\Big ( {1 \over N} \Big ).
\end{equation}
\end{proposition}

\noindent
{\rm Proof.} \quad Recalling (\ref{P1}), simple manipulation of (\ref{FZ1}) in the case
$\beta = 2$ gives
\begin{align}\label{Fe}
 Z_N(2) = &{1 \over N!}   \Big ( {1 \over 2 R} \Big )^{N } e^{-2 K_N}
\prod_{l=1}^N R^2 \int_0^{2 \pi} d \phi_l \int_0^\pi d \theta_l \,
 \Big ( \cos {\theta_l \over 2} \Big )^{2N-1 + 2qN}
 \Big ( \sin {\theta_l \over 2} \Big )^{1 + 2QN}  \nonumber \\
& \times \prod_{1 \le j < k \le N} \Big ( {v_j \over u_j} - {v_k \over u_k} \Big )
\Big ( {\bar{v}_j \over \bar{u}_j} - {\bar{v}_k \over \bar{u}_k} \Big ).
\end{align}
Making use of the Vandermonde determinant formula
$$
\prod_{1 \le j < k \le N}( x_k - x_j) = \det [ x_j^{k-1}]_{j,k=1,\dots,N},
$$
the readily verified orthogonality
$$
R^2 \int_0^{2 \pi} d \phi \int_0^\pi d \theta \, \sin \theta \, g(\theta) \Big ( {v \over u} \Big )^m
 \Big ( {\bar{v} \over \bar{u}} \Big )^n = 2 \pi R^2
  \delta_{m,n} \int_0^\pi g(\theta) \Big ( \tan {\theta \over 2} \Big )^{2n} \sin \theta \,
 d \theta
 $$
 valid for general $g$, and the Euler beta integral written in the form
 $$
 2 \int_0^{\pi/2} (\sin \theta)^{2a+1} (\cos \theta)^{2b+1}  \, d\theta =
 {\Gamma(a + 1) \Gamma(b+1) \over \Gamma(a+b+2)}
 $$
 the integral (\ref{Fe}) can be factorized into a product of one dimensional integrals with
 gamma function evaluations to give
 \begin{equation}\label{Ff}
  Z_N(2) = (2 \pi R)^{N}  e^{-2 K_N}
  \prod_{l=0}^{N-1} {\Gamma(l+NQ + 1) \Gamma(l+Nq + 1) \over
  \Gamma(N(1 + q + Q) + 1)}.
\end{equation}

A formula more immediately suited for asymptotic analysis can be obtained by introducing the
Barnes $G$-function.
This satisfies the functional equation $G(z+1) = \Gamma(z) G(z)$,
and can be given meaning for all complex $z$. In particular, it is known that for general
$\alpha$,
$$
\prod_{l=0}^{N-1} \Gamma(1 + l + \alpha) =
{G(N+\alpha + 1) \over G(\alpha + 1) }
$$
(see e.g.~\cite[eq.~(4.183)]{Fo10}) allowing (\ref{Ff}) to be rewritten
\begin{equation}\label{Ffa}
  Z_N(2) = (2 \pi R)^{N}  e^{-2 K_N}
 {1 \over ( \Gamma(N(1 + q + Q) + 1))^N}
 { G(N(1+Q)+1) G(N(1+q)+1) \over G(NQ + 1) G(Nq+1)}.
 \end{equation}
 In  (\ref{Ffa}), using Stirling's formula for the gamma function, the known asymptotic formula
 for the Barnes $G$-function
 $$
 \log G(x+1) \mathop{\sim}\limits_{x \to \infty}
 {x^2 \over 2} \log x - {3 \over 4} x^2 + {x \over 2} \log 2 \pi - {1 \over 12} \log x +
 \zeta'(-1) + {\rm O} \Big ( {1 \over x} \Big )
 $$
(see e.g.~\cite[eq.~(14) ]{We00}) and recalling the explicit form (\ref{C2}),
the stated expansion (\ref{Fd}) then follows.
\hfill $\square$\\

We remark that integrating both sides of
 (\ref{pz}) and using (\ref{Ff}) is an alternative way to deduce the normalization
 (\ref{C2}).

 \section{Correlation functions}\label{Section5}
 \setcounter{equation}{0}
 Throughout this section, we will work directly with the variables in the complex plane
 as implied by the eigenvalue problem, and are thus considering  the RHS of (\ref{pz})
 in the case $\beta = 2$.

 The general structure of the latter, being of the form
 \begin{equation}\label{hx}
 \prod_{l=1}^Nh(|z_l|)\prod_{1\leq j<k\leq N}|z_k-z_j|^2,\hspace{1cm}h(r):=\frac{r^{2QN}}{(1+r^2)^{(Q+q+1)N+1}}
 \end{equation}
 tells us that the $k$-point correlation function has the determinantal form
 \begin{equation}\label{Kx1}
 \rho_{(k)}(\mathbf{r}_1,\ldots, \mathbf{r}_k)={\rm det}[K(\mathbf{r}_\mu,\mathbf{r}_\gamma)]_{\mu,\gamma=1,\ldots,k},
 \end{equation}
 where the so-called correlation kernel $K$ is given by
 \begin{equation}\label{Hf}
 K(\mathbf{r}_\mu,\mathbf{r}_\gamma)=\frac{1}{\pi}(h(r_\mu)h(r_\gamma))^{1/2}H(r_\mu r_\gamma e^{i(\theta_\mu-\theta_\gamma)}), \hspace{1cm} H(z):=\frac{1}{2}\sum_{j=1}^N\frac{z^{j-1}}{\int_0^\infty h(r)r^{2j-1}dr}
 \end{equation}
 (here $(r,\theta)$ are the polar coordinates of $\mathbf{r}$). This follows from a simple calculation using the method of orthogonal polynomials (see e.g. \cite[Prop. 15.3.1]{Fo10}).

 We seek a form of $K$ suitable for asymptotic analysis.

 \begin{proposition}\label{p2p}
 Let
 \begin{equation}\label{Hf1a}
 \tilde{h}(r):=\frac{1}{(1+r^2)^{(Q+q+1)N+1}}
 \end{equation}
 and
 \begin{equation}\label{J}
 J(a,b;z):=\frac{1}{B(a,b)}\int_0^z \frac{t^{a-1}}{(1+t)^{b+a}}dt,
 \end{equation}
 where
 \begin{equation}\label{B1}
 B(a,b):=\int_0^\infty \frac{t^{a-1}}{(1+t)^{b+a}}dt= \frac{\Gamma(a)\Gamma(b)}{\Gamma(a+b)}.
 \end{equation}
 Furthermore, use (\ref{J}) to define
 \begin{equation}\label{HH}
 \tilde{H}(z)=(Q+q+1)N(1+z)^{(Q+q+1)N-1}(J(QN,(q+1)N;z)-J((Q+1)N,qN;z)).
 \end{equation}

 In terms of the quantities $\tilde{h}$ and $\tilde{H}$ we have
 \begin{equation}\label{Hf1}
 K(\mathbf{r}_\mu,\mathbf{r}_\gamma)=\frac{1}{\pi}(\tilde{h}(r_\mu)\tilde{h}(r_\gamma))^{1/2}\tilde{H}(r_\mu r_\gamma e^{i(\theta_\mu-\theta_\gamma)}),
 \end{equation}
 up to a factor which does not contribute to (\ref{Kx1}).
 \end{proposition}
 \noindent
{\rm Proof.} \quad Comparing (\ref{Hf1}) to (\ref{Hf}), we see that the task is to find a summation formula, by way of an integral representation, of the summation defining $H(z)$ in (\ref{Hf}). Straightforward working establishes that the latter satisfies the first order differential equation
\begin{equation}\label{Hb}
H'(z)+a(z)H(z)=b(z)
\end{equation}
where
\begin{align}
a(z)&=\frac{1}{1+z}\left( \frac{QN}{z}-(q+1)N+1\right), \label{aa}\\
b(z)&=\frac{QN\Gamma((Q+q+1)N+1)}{z(1+z)\Gamma(QN+1)\Gamma((q+1)N)}-\frac{z^{N-1}}{1+z}\frac{\Gamma((Q+q+1)N+1)}{\Gamma(N(Q+1))\Gamma(qN)}. \label{bb}
\end{align}

 According to the method of integrating factors, choosing $I(z)$ such that
 \begin{equation}\label{Ia}
 I'(z)=I(z)a(z)
 \end{equation}
 allows (\ref{Hb}) to be written
 $$
 \frac{d}{dz}(I(z)H(z))=b(z)I(z).
 $$
 Consequently $H(z)$ can be expressed in terms of $I(z)$ and $b(z)$ according to
 \begin{equation}\label{HC}
 H(z)=\frac{1}{I(z)}\int_0^zb(t)I(t)dt+C
 \end{equation}
 for some $C$ independent of $z$.

 Solving (\ref{Ia}) gives
 $$
 I(z)=\frac{z^{QN}}{(1+z)^{(Q+q+1)N-1}}.
 $$
 We substitute this and (\ref{bb})  into (\ref{HC}), then take  the limit $z\rightarrow 0$ to deduce that  $C=0$ and thus conclude
 \begin{align}
 H(z)&=\frac{(1+z)^{(Q+q+1)N-1}}{z^{QN}} \left( \frac{QN}{B(QN+1,(q+1)N)}\int_0^z\frac{t^{QN-1}}{(1+t)^{(Q+q+1)N}}dt \right. \notag \\
 &\left. \hspace{5cm}-\frac{qN}{B((Q+1)N,qN+1)}\int_0^z\frac{t^{(Q+1)N-1}}{(1+t)^{(Q+q+1)N}}dt \right) \label{Hx4}.
 \end{align}

 Use now of the recurrences
 $$
 B(x+1,y)=\frac{x}{x+y}B(x,y), \hspace{1cm}B(x,y+1)=\frac{y}{x+y}B(x,y)
 $$
 in (\ref{Hx4}) gives the form (\ref{HH}), but with an extra factor of $1/z^{QN}.$ This latter factor is essentially cancelled by the factor  of $r^{QN}$ in $(h(r))^{1/2}$ (recall (\ref{hx})) in the sense that with $\tilde{h}$ specified by (\ref{Hf1a}) and
 \begin{equation}\label{Hth}
 \tilde{H}(z)=z^{QN}H(z)
 \end{equation}
 we have that
 $$
 {\rm det}\left[(h(r_\mu) h(r_\gamma))^{1/2}H(r_\mu r_\gamma e^{i(\theta_\mu - \theta_\gamma)})\right]_{\mu,\gamma=1,\ldots,k}=
 {\rm det} \left[(\tilde{h}(r_\mu) \tilde{h}(r_\gamma))^{1/2}\tilde{H}(r_\mu r_\gamma e^{i(\theta_\mu - \theta_\gamma)})\right]_{\mu,\gamma=1,\ldots,k}.
 $$
 Since the above working shows that the formulas  (\ref{Hth}) and (\ref{HH}) for $\tilde{H}(z)$ are consistent, we have established (\ref{Hf1}).
 \hfill $\square$\\

 \subsection{Global scaling}\label{s5.1}
 In the variables of the RHS of (\ref{pz}), we know from (\ref{rr}) that the support of the
 underlying background charge density is between radii $\tilde{r}_Q$ and $\tilde{r}_q$,
 which are independent of $N$. Furthermore the uniform background on the sphere
 maps, under the stereographic projection, to the background in the plane
 \begin{equation}\label{bp}
- \rho_b(r) = - {N (1 + Q + q) \over \pi (1 + r^2)^2}.
\end{equation}

On the sphere, according to Proposition \ref{p1} the background density specified as the uniform value
$- \rho_b$ within the spherical annulus and zero density outside is the solution of the
integral equation
$$
- \int_S \rho((\theta,\phi)) \log | u' v - u v'| \, dS =
C_N - NQ \log \sin {\theta' \over 2} - N q \cos {\theta' \over 2}.
$$
As such $\rho_b$ provides the minimum of the energy functional
\begin{align*}
E[\rho] = & - \int_S \rho((\theta, \phi)) \Big ( N Q \log \sin {\theta \over 2} +
N q \log \cos {\theta \over 2} \Big ) \, dS \nonumber \\
& - {1 \over 2} \int_S dS_1  \,  \rho((\theta_1, \phi_1)) \int_S dS_2 \,   \rho((\theta_2, \phi_2))
\log |u_2 v_1 - u_1 v_2|.
\end{align*}
On the other hand, we know that to leading order the density of the mobile particles in the
plasma can be characterised by minimizing this same energy functional (see
e.g.~\cite{BH09} and references therein). Thus to leading order it must be that that
the particle density is equal to $\rho_b$. When projected to the plane, this means that to
leading order the particle density will be confined between  radii $\tilde{r}_Q$ and $\tilde{r}_q$,
and will have profile given by (\ref{bp}) (without the minus signs).
We will see that this prediction is confirmed by explicit calculation, and we will show too
that the correction terms are exponentially small in $N$.

According to Proposition \ref{p2p}
\begin{equation}\label{r1}
\rho_{(1)}(\mathbf{r})=\frac{(Q+q+1)N}{\pi(1+r^2)^2}\Big(J(QN,(q+1)N;r^2)-J((Q+1)N,qN;r^2) \Big).
\end{equation}
Our task is to compute the large $N$ asymptotic form of this expression.

\begin{proposition}
For asymptotically large values of $N$ the density (\ref{r1}) vanishes outside the annulus $r\in [{r}_Q, {r}_q]$ up to exponentially small terms in $N$, while inside this annulus, again up to exponentially small terms $\rho_{(1)}(r)=\rho_b(r)$ as specified by (\ref{hx}).
\end{proposition}
\noindent
{\rm Proof.} \quad
According to (\ref{r1}) we require the large $N$ form of $J(\alpha N,\beta N;x)$ for $x>0$ fixed and $\alpha,\beta>0$. From the definition (\ref{J}) we see that the $N$-dependent portion of the integrand in the definition of $J(\alpha N,\beta N;x)$ can be written
\begin{equation}\label{ee}
e^{N(\alpha \log t-(\alpha+\beta)\log (1+t))}.
\end{equation}
This has a single maximum at $t=t_0:=\alpha/\beta$, and correspondingly  $J(\alpha N,\beta N;x)$ is exponentially small when $t_0$
 is not part of the range of integration. Consequently, up to exponentially small terms in $N$
 \begin{equation}\label{g3}
 J(\alpha N, \beta N;x)\sim
\left\{
\begin{tabular}{ll}
$1,$ & $x>\alpha/\beta$ \\
$0,$ & $x<\alpha/\beta.$
\end{tabular}
\right.
\end{equation}
The stated result now follows by using this result in (\ref{r1}).
\hfill $\square$\\

We now turn our attention to the large $N$ behaviour of the truncated two-point correlation function,
\begin{equation}\label{rhot}
\rho_{(2)}^T(\mathbf{r}_1,\mathbf{r}_2):=\rho_{(2)}(\mathbf{r}_1,\mathbf{r}_2)-\rho_{(1)}(\mathbf{r}_1)\rho_{(1)}(\mathbf{r}_2).
\end{equation}
According to (\ref{Kx1}) and (\ref{Hf}) (for later purposes this is more useful than (\ref{Hf1})) this has the explicit form
\begin{equation}\label{r2}
\rho_{(2)}^T(\mathbf{r}_1,\mathbf{r}_2)=-\frac{1}{\pi^2}h(\mathbf{r}_1)h(\mathbf{r}_2)|H(r_1 r_2 e^{i(\theta_1-\theta_2)})|^2.
\end{equation}
For $\mathbf{r}_1\not=\mathbf{r}_2$ and fixed as $N\rightarrow \infty$, on the scale of the spacing between eigenvalues the eigenvalues at $\mathbf{r}_1$ and $\mathbf{r}_2$ are effectively an infinite distance apart. They will thus be uncorrelated and so we expect $\rho_{(2)}^T(\mathbf{r}_1,\mathbf{r}_2)\rightarrow 0$ as $N\rightarrow \infty$. On the other hand we now have that with $\mathbf{r}_1=\mathbf{r}_2$ the truncated two-particle correlation is equal to $-(\rho_{(1)}(\mathbf{r}_1))^2$ which we know is proportional to $N^2$ for $\mathbf{r}_1$ inside the annulus.

To quantify this behaviour, consideration of fluctuation formulas for linear statistics (see Section \ref{s5.2} below) suggests that the appropriate quantity to analyze is
$$
I[a]:=\int_{\mathbb{R}^2} d\mathbf{r}_1 a(\mathbf{r}_1)  \int_{\mathbb{R}^2}d\mathbf{r}_2 a(\mathbf{r}_2)  \Big(\rho_{(2)}^T(\mathbf{r}_1,\mathbf{r}_2) +\delta(\mathbf{r}_1-\mathbf{r}_2)\rho_{(1)}(\mathbf{r}_1) \Big),
$$
for all $a(\mathbf{r})$ sufficiently smooth.

\begin{proposition}\label{p5.3}
Let $a(\mathbf r) = a(r) = \alpha(r^2)$ so that $a(\mathbf r)$ is rotationally invariant, and with $r^2 = s$, let $\alpha(s)$ be twice differentiable with respect to $s$ We have
\begin{equation}\label{y5}
\lim_{N \rightarrow \infty}I[a]=\int_{Q/(1+q)}^{(Q+1)/q} (\alpha' (s))^2 ds.
\end{equation}
\end{proposition}

\noindent
{\rm Proof.} \quad
We see from (\ref{Hf}) that
$$
\int_0^{2\pi} d \theta_1 \int_0^{2\pi} d \theta_2 |H(r_1 r_2 e^{i(\theta_1-\theta_2)})|^2=\pi^2\sum_{j=1}^N \frac{(r_1r_2)^{2(j-1)}}{(\int_0^\infty h(r) r^{2j-1}dr)^2}.
$$
Consequently, with $a(\mathbf{r})=\alpha(r^2)$,
\begin{equation}\label{g6}
I[a]=-\sum_{j=1}^N\left(\frac{\int_0^\infty \alpha(r^2)h(r)r^{2j-1} dr}{\int_0^\infty h(r)r^{2j-1} dr} \right)^2+\sum_{j=1}^N \frac{\int_0^\infty (\alpha(r^2))^2h(r)r^{2j-1} dr}{\int_0^\infty h(r)r^{2j-1} dr}.
\end{equation}
Writing $j=Nt$, $t:=(j-1)/N$, and thus $0\leq t<1$, the large $N$ form of the integrals can be determined as in the proof of Proposition \ref{p2p}. In particular, after changing variables $s=r^2$, the maximum of the $N$-dependent factor of the integrands (i.e. $h(r)r^{2j-2}$) is seen to occur at
$$
s=s_0(t)=\frac{Q+t}{q+1-t}.
$$
Thus we expand
\begin{align*}
&h(r)r^{2(j-1)}\Big|_{r^2=s}\sim h(\sqrt{s_0})s_0^{(j-1)}e^{-N(s-s_0)^2/2\sigma^2},\hspace{1cm}\sigma^2=\frac{(Q+t)(q+Q+1)}{(q+1-t)^3}\\
&\hspace{2cm} \alpha(s)\sim \alpha(s_0)+(s-s_0)\alpha'(s_0)+\frac{1}{2}(s-s_0)^2\alpha''(s_0).
\end{align*}
Substituting in (\ref{g6}) allows us to conclude that for large $N$
\begin{align*}
I[a]&\sim\sum_{j=1}^N(\alpha'(s_0))^2\frac{\int_0^\infty (s-s_0)^2 e^{N(s-s_0)^2/2\sigma^2}ds}{\int_0^\infty e^{-N(s-s_0)^2/2\sigma^2}ds} \\
&=\frac{(q+Q+1)}{N}\sum_{j=1}^N(\alpha'(s_0))^2\frac{(Q+t)}{(q+1+t)^3}.
\end{align*}
But this last expression is just the Riemann sum approximation to an integral. After changing variables, (\ref{y5}) results.
\hfill $\square$\\

We remark that for $\alpha(s)$ as in (\ref{p5.3})
\begin{align}\label{D}
\int_{Q/(1+q)}^{(Q+1)/q}(\alpha'(s))^2ds&=\frac{1}{2}\int_{{r}_Q}^{{r}_q}\Big( \frac{1}{r}\frac{d}{dr}\Big)^2 a(r) dr \nonumber \\
&=\frac{1}{4\pi} \int_{D_{[{r}_Q, {r}_q]}}\Big( \frac{\partial ^2}{\partial x^2}+ \frac{\partial ^2}{\partial y^2} \Big)a(\mathbf{r})d\mathbf{r}
\end{align}
where $D_{[{r}_Q, {r}_q]}$ denotes the annulus with inner radius ${r}_Q$ and outer radius ${r}_q$. This is consistent with the expected large $N$ form \cite{Ja95}
$$
\rho_{(2)}^T(\mathbf{r}_1,\mathbf{r}_2)+\delta(\mathbf{r}_1-\mathbf{r}_2)\rho_{(1)}(\mathbf{r}_1)=\bigtriangledown_{\mathbf{r}_1}^2\delta(\mathbf{r}_1-\mathbf{r}_2)
$$
for $\mathbf{r}_1$ and $\mathbf{r}_2$ away from the boundary of the annulus.

Coulomb gas theory predicts very different behaviour for $\mathbf{r}_1$, $\mathbf{r}_2$ within the boundary layer of the support \cite{Ja95}. Consider for definiteness the inner edge. The theory of \cite{Ja95} predicts
\begin{equation}\label{ag}
\lim_{N\rightarrow \infty}\Big( \frac{{r}_Q^2}{\rho_{b}(\mathbf{r}_Q)}\Big)\rho_{(2)}^T\Big((\mathbf{r}_Q+\frac{s_1}{\sqrt{\rho_b(\mathbf{r}_Q)}},\theta_1),(\mathbf{r}_Q+\frac{s_2}{\sqrt{\rho_b(\mathbf{r}_Q)}},\theta_2) \Big) =-\frac{g(s_1, s_2)}{4\pi^2 |1-e^{i(\theta_1-\theta_2)}|^2}
\end{equation}
where $g(s_1,s_2)$ has the property that
$$
\int_{-\infty}^\infty ds_1 \int_{-\infty}^\infty ds_2 \,g(s_1,s_2)=1.
$$
This result has previously been exhibited for the one-component plasma at $\beta=2$ in the case of disk geometry \cite{CPR87}, as has the analogue of (\ref{ag}) for the same system but now in an ellipse geometry \cite{FJ96} (the latter is equivalent to the partially symmetric Ginibre ensemble of
complex random matrices \cite{FKS97}). It can readily be checked in the present setting of a projected spherical annulus.

\begin{proposition}
The limit formula (\ref{ag}) holds true with
\begin{equation}\label{gs}
g(s_1,s_2)=\frac{2}{\pi}e^{-2s_1^2-2s_2^2}.
\end{equation}
\end{proposition}
\noindent
{\rm Proof.} \quad
Our main tool is an asymptotic formula for $J(\alpha N,\beta N;x)$ valid for $x$ bounded away from the real axis. In this case, along a ray from the origin to $x$, the corresponding integrand oscillates rapidly and the main contribution to the integral comes from the neighbourhood of the end point at $x$. To determine the latter we follow a strategy used on the incomplete gamma function in \cite{CPR87}, involving a particular integration by parts.

The integration by parts in turn is initiated by writing the integrand in terms of a derivative of its own functional form,
$$
\frac{t^{\alpha N-1}}{(1+t)^{(\alpha+\beta)N}}=\frac{1+t}{N(\alpha-\beta t)} \frac{d}{dt}\Big(\frac{t^{\alpha N}}{(1+t)^{(\alpha+\beta)N}} \Big).
$$
With $x$ bounded away from the real axis, substitution of this in the definition (\ref{J}) and integration by parts shows
\begin{equation}\label{Jsa}
J(\alpha N, \beta N; x)\sim \frac{1}{B(\alpha N,\beta N)}\frac{1}{N(\alpha -\beta x)}\frac{x^{\alpha N}}{(1+x)^{(\alpha+\beta)N}}\Big(1+ O\Big( \frac{1}{N}\Big) \Big).
\end{equation}
Furthermore, with $\gamma:=\alpha/\beta$, use of Stirling's formula shows
\begin{equation}\label{Js1}
B(\alpha N,\beta N)\sim \sqrt{\frac{2\pi}{\alpha N}}\Big( \frac{\gamma}{1+\gamma}\Big)^{\alpha N}\Big( \frac{1}{1+\gamma}\Big)^{\beta N+1/2}.
\end{equation}

Substituting (\ref{Js1}) in (\ref{Jsa}), then substituting the result in (\ref{Hx4}) and recalling (\ref{r2}) shows
\begin{align}\label{Js2}
\rho_{(2)}^T(z_1,z_2)\sim& -\frac{N(Q+q+1)^2}{2 \pi^3 Q}\frac{|z_1 z_2|^{2QN}}{((1+|z_1|^2)(1+|z_2|^2))^{(Q+q+1)N+1}}\notag\\
&\times \Big(\frac{1+ {r}_Q^2}{{r}_Q^2} \Big)^{2QN}(1+ {r}_Q^2)^{2(q+1)N-1}\Big| \frac{1}{1-z_1\overline{z}_2/{r}_Q^2}\Big|^2.
\end{align}
Next, we must substitute for $z_1$ and $z_2$ as required by the LHS of (\ref{ag}). Appropriate large-$N$ expansion of the resulting terms on the RHS of (\ref{Js2}) gives the RHS of (\ref{ag}) with $g(s_1,s_2)$ therein given by (\ref{gs}).
\hfill $\square$

 \subsection{Local scaling}\label{s5.2}

 A feature of the global scaling of the previous section is that the area of the annulus remains fixed as the number of eigenvalues tends to infinity. In contrast, we know from Section \ref{Section4} that the thermodynamic limit is such that the volume of the annulus tends to infinity while the density of the eigenvalues stays fixed. At the level of the correlation functions, due to the scale invariance of the logarithmic potential, the thermodynamic limit is equivalent to a local scaling in which the position variables are measured on the scale of the (linear) inter-particle spacing. This can be achieved by rewriting each polar coordinate $(r,\theta)$ in terms of a cartesian coordinate $(x,y)$ according to
 \begin{equation}\label{rxy}
 r=X+\frac{x}{\sqrt{\rho_b(X)}}, \hspace{1cm} \theta=\frac{y}{X\sqrt{\rho_b(X)}}
 \end{equation}
 for $X\in({r}_Q, {r}_q)$, where $\rho_b(X)$ is given by (\ref{bp}). We seek the asymptotic form of the correlation kernel (\ref{Hf}) under this scaling.

 \begin{proposition}\label{p4}
 Let the polar coordinates of $\mathbf{r}_\mu$, $\mathbf{r}_\gamma$ be replaced by the scaled cartesian coordinates (\ref{rxy}). The correlation kernel (\ref{Hf})
 \begin{align}\label{K1a}
 \frac{1}{\rho_b(X)}K(\mathbf{r}_\mu,\mathbf{r}_\gamma)\sim& e^{iNX(y_\mu-y_\gamma)/(\sqrt{p_0}(1+X^2))}\notag \\
 & \times{\rm exp}\Big( -\frac{\pi}{2}(x_\mu-x_\gamma)^2 -\frac{\pi}{2}(y_\mu-y_\gamma)^2+i\pi(x_\mu+x_\gamma)(y_\mu-y_\gamma)+O\Big( \frac{1}{N^{1/2}} \Big) \Big)
 \end{align}
 and thus, up to terms $O(1/N^{1/2})$
 \begin{equation}\label{Kb}
 \Big( \frac{1}{\rho_b(X)}\Big)^k\rho_{(k)}(\mathbf{r}_1,\ldots, \mathbf{r}_k)\sim{\rm det}[e^{-\frac{1}{2}(x_\mu-x_\gamma)^2-\frac{1}{2}(y_\mu-y_\gamma)^2+i(x_\gamma y_\mu-x_\mu y_\gamma)}]_{\mu,\gamma=1,\ldots,k}.
 \end{equation}

 \end{proposition}

 \noindent
{\rm Proof.} \quad
 We recall that $K(\mathbf{r}_\mu,\mathbf{r}_\gamma)$ is given in terms of $\tilde{h}$ and $\tilde{H}$ according to (\ref{Hf1}). But it follows from (\ref{g3}) that for $\mathbf{r}_\mu, \mathbf{r}_\gamma$ in the annulus and within $O(1/\sqrt{N})$ of the real axis
 \begin{equation}\label{Ke2}
 \tilde{H}(r_\mu r_\gamma e^{i(\theta_\mu-\theta_\gamma)})\sim (Q+q+1)N(1+r_\mu r_\gamma e^{i(\theta_\mu-\theta_\gamma)})^{(Q+q+1)N-1}
 \end{equation}
 up to terms $O(1\sqrt{N})$. Recalling now the definition (\ref{Hf1a}) of $\tilde{h}$ we thus have
 $$
 \frac{1}{\rho_b(X)}K(\mathbf{r}_\mu, \mathbf{r}_\gamma)\sim\Big( \frac{1+r_\mu r_\gamma e^{i(\theta_\mu-\theta_\gamma)}}{(1+r_\mu^2)^{1/2}(1+r_\gamma^2)^{1/2}}\Big)^{(Q+q+1)N-1}
 $$
 up to terms $O(1/\sqrt{N})$. Introducing (\ref{rxy}), the form (\ref{K1a}) now follows upon elementary computation. And this used in (\ref{Kx1}), after observing that the first exponential factor on the RHS does not contribute to the determinant, nor does the factors $e^{i(x_\mu y_\mu- x_\nu y_\nu)}$, implies (\ref{Kb}).
 \hfill $\square$\\

 We would expect that the correlations in this bulk scaling limit would be independent of the geometry, and thus be the same as for the disk geometry for example. With $z=x+iy$ the latter are given by \cite[Prop. 15.3.2]{Fo10}
 $$
 {\rm det}[e^{-\pi(|z_\mu|^2+|z_\gamma|^2)/2}e^{z_\mu\overline{z}_\gamma}]_{\mu,\gamma=1,\ldots,k}
 $$
 which is indeed identical to the RHS of (\ref{Kb}).

 In (\ref{rxy}) we required that $X$ be strictly inside the annulus. With this assumption we were able to make use of (\ref{g3}). A physically different regime is to scale coordinates to have $O(1)$ spacing in the neighbourhood of a boundary of the annulus (for definiteness this will be taken to be the inner boundary). With $|z|^2= {r}_Q+O(1/\sqrt{N})$ the function $J$ in (\ref{g3}) exhibits a crossover function form linking the two limiting values exhibited in (\ref{g3}).

 \begin{proposition}
 Let $J$ be specified by (\ref{J}). We have
 \begin{equation}\label{Js}
 \lim_{N\rightarrow \infty} J\Big(\alpha N,\beta N;\frac{\alpha}{\beta}+\frac{X}{\sqrt{NC_0}}\Big)=\frac{1}{2}+\frac{1}{2}{\rm erf}\Big(\frac{X}{\sqrt{2}}\Big)
 \end{equation}
 where ${\rm erf}(x):=\frac{2}{\sqrt{\pi}}\int_0^xe^{-s^2}ds$ denotes the error function and
 \begin{equation}\label{Jc}
 C_0=\frac{\beta^3}{\alpha(\alpha+\beta)}.
 \end{equation}
 \end{proposition}

  \noindent
{\rm Proof.} \quad
With $C_0$ as in (\ref{Jc}), we see that expanding the exponent on the RHS of (\ref{ee}) about its maximum at $t=t_0$ to second order gives
$$
\frac{t^{\alpha N}}{(1+t)^{(\alpha+\beta)N}}\sim\frac{t^{\alpha N}}{(1+t_0)^{(\alpha+\beta)N}}e^{-NC_0(t-t_0)^2/2}.
$$
Recalling the definition (\ref{J}) of $J$, it follows that
$$
J\Big(\alpha N,\beta N;\frac{\alpha}{\beta}+\frac{X}{\sqrt{NC_0}}\Big)\sim \sqrt{\frac{2}{\pi}}\int_{-\infty}^Xe^{-t^2/2}dt
$$
which implies (\ref{Js}).
\hfill $\square$\\

Writing
\begin{equation}\label{zr}
z=r_Q+\frac{X}{\sqrt{\rho_b(r_Q)}}+i\frac{Y}{\sqrt{\rho_b(r_Q)}}
\end{equation}
so that the (complex) coordinate is scaled and centred about the inner boundary, it follows from (\ref{Js}) that
\begin{equation}\label{u5}
\lim_{N\rightarrow \infty} J(QN,(q+1)N;z_\mu \overline{z}_\gamma)=\frac{1}{2}+\frac{1}{2}{\rm erf}\Big(\frac{X_\mu+X_\gamma+i(Y_\mu-Y_\gamma)}{\sqrt{2}} \Big).
\end{equation}

\begin{proposition}
Let $\mathbf{r}_1,\ldots,\mathbf{r}_k$ be centred and scaled about the inner boundary of the annulus as implied by (\ref{zr}). We have
\begin{equation}\label{u6}
\lim_{N\rightarrow \infty}\Big(\frac{1}{\rho_b(r_Q)} \Big)^{k}\rho_{(k)}(\mathbf{r}_1,\ldots,\mathbf{r}_k)={\rm det}[H((X_\mu,Y_\mu),(X_\gamma,Y_\gamma))]_{\mu,\gamma=1,\ldots,k}
\end{equation}
where
\begin{equation}\label{Hes}
H((X_\mu,Y_\mu),(X_\gamma,Y_\gamma))=e^{-\frac{1}{2}(X_\mu-X_\gamma)^2-\frac{1}{2}(Y_\mu-Y_\gamma)^2+i(X_\gamma Y_\mu-X_\mu Y_\gamma)}\Big(\frac{1}{2}+\frac{1}{2}{\rm erf}\Big(\frac{X_\mu+X_\gamma+i(Y_\mu-Y_\gamma)}{\sqrt{2}} \Big)\Big).
\end{equation}
\end{proposition}

  \noindent
{\rm Proof.} \quad
 The only difference between this and the proof of Proposition \ref{p4} is that the function $J$ which implicitly multiplies the RHS of (\ref{Ke2}) is no longer unity but rather is given by (\ref{u5}). Hence the only difference between (\ref{u6}) and (\ref{Kb}) is this extra factor.
 \hfill $\square$\\

 The scaled edge correlation function (\ref{u6}) is precisely the same as found for the scaled edge correlation in the Ginibre ensemble of complex Gaussian matrices \cite{FH98}, which in turn is equivalent to the one-component plasma in an annulus at $\beta=2$ with soft wall boundary conditions.

 \subsection{Fluctuation formulas for linear statistics}
 Knowledge of the one and two point correlation functions in the global scaling regime
 gives information on the
 mean and variance of a linear statistic. The latter is specified as the random variable
 $A = \sum_{j=1}^N a(z_j)$ where $\{z_j\}$ are the eigenvalues of the random matrix.
 Thus we have
 \begin{equation}\label{Me}
 \langle  A  \rangle = \int_{\mathbb R^2} \rho_{(1)}(z) a(z) \, dx dy
 \end{equation}
 and
\begin{equation}\label{Var}
{\rm Var} \, A =   \int_{\mathbb R^2} dx_1 dy_1 \, a(z_1)  \int_{\mathbb R^2} dx_2 dy_2 \, a(z_2)
\Big ( \rho_{(2)}^T(z_1,z_2) +  \delta(z_1-z_2)  \rho_{(1)}(z) \Big )
 \end{equation}
 (for the latter equation, see e.g.~\cite[Prop.~14.3.2]{Fo10}), where
 $\rho_{(2)}^T(z_1,z_2)$ is specified by (\ref{rhot}).
 As previously remarked, the global scaling regime corresponds from a statistical mechanics
 viewpoint to an infinite density limit, since the $N$ eigenvalues are confined to an annulus of
 fixed radius as $N$ increases to infinity. It is precisely this limit (see
 e.g.~\cite[Ch.~14]{Fo10}) that gives rise to universal behaviour by way of a Gaussian
 fluctuation formula, with a variance which is ${\rm O}(1)$.

 We will consider first the limiting form of the mean and variance. For the mean,
 it follows from (\ref{bp}) and (\ref{rr}) that
 \begin{equation}\label{Me1}
 \lim_{N \to \infty} {1 \over N}  \langle  A  \rangle = {(1 + Q + q) \over \pi}
 \int_{D_{[\tilde{r}_Q,\tilde{r}_q]}} {a(z) \over (1 + |z|^2)^2} \, dx dy,
 \end{equation}
 where $D_{[\tilde{r}_Q, \tilde{r}_q]}$ is defined as in (\ref{D}). In the case that
 $a(z) = a(|z|)$ so that the linear statistic is rotationally invariant, we have from (\ref{y5})
 with $\alpha(r^2) = a(r)$ that the limit variance is given by
 \begin{equation}\label{y5a}
  \lim_{N \to \infty}   {\rm Var} \, A =   \int_{(Q+1)/q}^{Q/(q+1)} (a'(\sqrt{s}))^2 \, ds
  \end{equation}

In this latter case the full distribution of $A$ can easily be obtained via an explicit calculation.
 This is analogous to the situation for the complex Ginibre ensemble \cite{Fo99}.

 \begin{proposition}
 Let $\langle \cdot \rangle$ denote the average with respect to the PDF corresponding to the
 RHS of (\ref{pz}) in the case $\beta = 2$. Let $a(z) = a(|z|)$ have a continuous derivative with respect to
 $r = |z|$ for $r$ in $D_{[\tilde{r}_q,\tilde{r}_Q]}$. For large $N$ we have that
 \begin{align}
 \Big \langle e^{i k \sum_{l=1}^N a(z_l)} \Big \rangle& \sim
 \exp \Big ( i k N(q+Q+1)
\int_{(Q+1)/q}^{Q/(q+1)} {a(\sqrt{s}) \over (1+s)^2} \, ds
- {k^2 \over 2} \int_{(Q+1)/q}^{Q/(q+1)} (a'(\sqrt{s}))^2 \, ds \Big ) \nonumber \\
& = \exp \Big ( i kN \lim_{N \to \infty} {1 \over N} \langle A \rangle - {k^2 \over 2}
\lim_{N \to \infty} {\rm Var} \, A \Big ).
\end{align}
Consequently, as $N \to \infty$, $A - \langle A \rangle$ is distributed as a standard Gaussian
with variance $\lim_{N \to \infty} {\rm Var} \, A$.
\end{proposition}

\noindent Proof. \quad Using an analogous integration procedure to that used in deriving
(\ref{Ff}), now using polar cordinates as in the workings of Sections
\ref{s5.1} and \ref{s5.2}, we readily obtain
\begin{equation}\label{5.5}
 \Big \langle e^{i k \sum_{l=1}^N a(z_l)} \Big \rangle
=
\prod_{j=1}^N {\displaystyle
\int_0^\infty e^{i k  a(\sqrt{s})}   \Big ( {s \over 1 + s} \Big )^{N Q}
\Big ( {1 \over 1 + s} \Big )^{N(q+1) + 1} s^{j-1} \, ds \over \displaystyle
\int_0^\infty   \Big ( {s \over 1 + s} \Big )^{N Q}
\Big ( {1 \over 1 + s} \Big )^{N(q+1) + 1} s^{j-1} \, ds }.
\end{equation}

We set $j=N t$, where $t:=j/N$ and thus $0 < t < 1$. Now writing the $N$ dependent terms in
the integrand in the exponential form
\begin{equation}\label{Mq}
\exp \Big ( N \Big [ Q \log {s \over 1 + s} +  (q+1) \log {1 \over 1 + s} +  t \log s \Big ]  \Big ),
\end{equation}
we see that the maximum occurs for
\begin{equation}\label{Mq1}
s = s_0(t) := { Q + t \over q + 1 - t}.
\end{equation}
Expanding (\ref{Mq}) to second order about $s_0(t)$, and expanding the factor $e^{i k a(s)}$
in the integrand of the numerator of (\ref{5.5}) to first order about $s = s_0(t)$ then completing
the square, we see that
$$
 \Big \langle e^{i k \sum_{l=1}^N \alpha(z_l)} \Big \rangle  \sim
 \exp \Big ( i k \sum_{j=1}^N a(\sqrt{s_0(t)}) - {k^2(Q+q+1) \over 2N}
 \sum_{j=1}^N {(Q + t) \over (q + 1 - t)^3 }(a'(\sqrt{s_0(t)}))^2 \Big ).
 $$
 Recalling now the definition of $t$ above (\ref{Mq}) we see that the sums are to leading
 order Riemann integrals. After a change of variables, the first line of
 (\ref{5.5}) results.  The second line follows  by using (\ref{Me1})
 and (\ref{y5}). \hfill $\square$\\

For linear statistics not rotationally invariant, there will be a contribution to the variance due to the universal
form of the surface correlations (\ref{ag}) \cite{CPR87,Fo99}. In the case that $a(\vec{r})$ is sufficiently smooth,
a proof of its explicit form, together with a proof of the corresponding Gaussian fluctuation formula, follows from
a more general theorem of Ameur, Hedenmalm and Makarov \cite{AHM08}. This latter theorem also includes
the setting of non-rotationally invariant linear statistics for the complex Ginibre ensemble, first established by
Rider and Vir\'ag \cite{RV06}. In the case of linear statistics dependent only on the angle (and thus not
smooth at the origin), the variance is typically no longer of order one, but nonetheless Gaussian fluctuation formulas
can still be established \cite{ER11}.

\subsection*{Acknowledgements}
This work was supported by the Australian Research Council. JF acknowledges financial support from the Eileen Colyers prize given by the department of mathematical sciences, Queen Mary University of London, as well as the generous hospitality at the University of Melbourne. The assistance in producing the figure of Wendy Baratta and Anthony Mays is acknowledged.


\providecommand{\bysame}{\leavevmode\hbox to3em{\hrulefill}\thinspace}
\providecommand{\MR}{\relax\ifhmode\unskip\space\fi MR }
\providecommand{\MRhref}[2]{%
  \href{http://www.ams.org/mathscinet-getitem?mr=#1}{#2}
}
\providecommand{\href}[2]{#2}

\end{document}